\documentclass[aps,
reprint,
superscriptaddress,
graphicx,
longbibliography]{revtex4-1}
\usepackage{blindtext}
\usepackage{braket}
\usepackage{graphicx} 
\usepackage[T1]{fontenc}
\usepackage{amsmath,amsthm,amsfonts}
\usepackage{bm}
\usepackage{varioref}
\usepackage[colorlinks=true,linkcolor=blue,citecolor=blue]{hyperref}
\usepackage{cleveref}
\title{Sections and Chapters}

\newcommand{\e}{\mathrm{e}}
\newcommand{\diag}{\mathrm{diag}}
\newcommand{\Tr}{\mathrm{Tr}}
\renewcommand{\Re}{\mathrm{Re}}
\renewcommand{\Im}{\mathrm{Im}}
\renewcommand{\i}{\mathrm{i}}

\begin{document}
\title{Heisenberg-scaling precision in the estimation of two parameters in a Mach-Zehnder interferometer}
\author{Atmadev Rai}
\affiliation{School of Mathematics and Physics, University of Portsmouth, Portsmouth PO1 3QL, UK}
\author{Danilo Triggiani}
\affiliation{School of Mathematics and Physics, University of Portsmouth, Portsmouth PO1 3QL, UK}
\affiliation{Dipartimento Interateneo di Fisica, Politecnico di Bari, Bari 70126, Italy}
\affiliation{INFN, Sezione di Bari, I-70126 Bari, Italy}
\author{Paolo Facchi}
\affiliation{INFN, Sezione di Bari, I-70126 Bari, Italy}
\affiliation{Dipartimento di Fisica, Universit\`{a} di Bari, I-70126 Bari, Italy}
\author{Vincenzo Tamma}
\email[]{vincenzo.tamma@port.ac.uk}
\affiliation{School of Mathematics and Physics, University of Portsmouth, Portsmouth PO1 3QL, UK}
\affiliation{Institute of Cosmology and Gravitation, University of Portsmouth, Portsmouth PO1 3FX, UK}
\begin{abstract}
Achieving the ultimate quantum precision in the estimation of multiple physical parameters simultaneously is a challenge in quantum metrology due to fundamental limitations and experimental challenges in harnessing the necessary quantum resources. We propose an experimentally feasible scheme to reach Heisenberg-scaling precision in the simultaneous estimation of two unknown phase parameters in a Mach-Zehnder interferometer by using a squeezed and a coherent state of light as input and homodyne detections at the outputs. Our results open a new research paradigm in harnessing the full quantum advantage in multiparameter estimation in distributed quantum sensing and quantum information processing, as enabled by multiphoton quantum interference with scalable experimental resources.
\end{abstract}

\maketitle
\section{Introduction}
Quantum metrology can enhance the sensitivity in the estimation of unknown physical parameters. In particular, when we employ classical light sources, sensitivity typically attains the shot-noise limit (SNL) of precision, with a scaling of $1/\sqrt{N}$, where $N$ is the average number of photons in the probe. On the other hand, utilizing quantum phenomena such as entanglement, quantum coherence, and squeezing, allows one to surpass the SNL and approach the Heisenberg limit, achieving a scaling of $1/N$~\cite{dowling2015quantum, zhou2018achieving, qian2019heisenberg,PhysRevA.33.4033}. In the last decade, several protocols have been proposed to achieve Heisenberg scaling for the estimation of a single parameter~\cite{PhysRevD.23.1693, PhysRevLett.100.073601, PhysRevLett.111.173601, Gramegna_2021, PhysRevResearch.3.013152, PhysRevA.91.032103, PhysRevLett.121.043604, triggiani2022estimation, PhysRevA.105.012607, PhysRevA.85.011801}. Different quantum states are used to enhance the sensitivity including squeezed sources, which have garnered attention due to their advantageous characteristics in experimental implementation and inherent quantum properties. Recently experimental advancements have reaffirmed the quantum metrological benefits provided by squeezing resources, thus playing a pivotal role in quantum metrology~\cite{PhysRevA.104.062603, PhysRevA.105.012607, lawrie2019quantum, pradyumna2020twin, aasi2013enhanced, PhysRevLett.88.231102, PhysRevA.79.033834, matsubara2019optimal, PhysRevA.85.010101, PhysRevLett.130.123603, PhysRevLett.130.070801}.

In recent years, the estimation of multiple physical parameters has attracted significant attention~\cite{PhysRevLett.111.070403, yue2014quantum, PhysRevA.102.022602, PhysRevA.90.062113, hong2021quantum, PhysRevA.94.042342, PhysRevLett.119.130504, PhysRevA.94.062312}, as it is preferable in terms of resources and technological applications to estimate multiple parameters simultaneously, rather than individually and at different times. Multiparameter estimation plays a pivotal role in various fields, including quantum imaging~\cite{kolobov2007quantum, spagnolo2012quantum, Genovese_2016}, biological system measurement~\cite{taylor2013biological, taylor2016quantum, mauranyapin2017evanescent}, astronomy~\cite{PhysRevA.95.063847, PhysRevA.96.062107}, sensor networks~\cite{komar2014quantum, nokkala2018reconfigurable}, quantum process tomography~\cite{zhou2015quantum, hayashi2006parallel}, and the estimation of gravitational wave parameters~\cite{freise2009triple, schnabel2010quantum}. Moreover, it finds applications in detecting inhomogeneous forces and gradients~\cite{koschorreck2011high, baumgratz2016quantum, PhysRevA.97.053603}. These diverse tasks are beyond single-parameter estimation.

Despite the broad range of applications and recent advancements in multiparameter estimation, significant challenges persist. For example, reliance on sources such as photon number states and entangled states poses challenges in their generation and the maintenance of quantum coherence~\cite{hong2021quantum, yao2022two}. Moreover, determining optimal measurement schemes capable of achieving Heisenberg-scaling sensitivity across multiple parameters simultaneously remains a pressing concern~\cite{polino2020photonic}. Additionally, some estimation techniques impose constraints on the values of unknown parameters, rendering them unsuitable for encoding within arbitrarily distributed networks~\cite{guo2020distributed,zhuang2020distributed,  PhysRevX.9.041023, PhysRevResearch.1.032024}. To the best of our knowledge, no scheme has been developed for achieving multiparameter estimation with Heisenberg scaling using scalable resources, such as squeezed light sources, and experimentally feasible measurements, such as homodyne measurements.

In this work, we consider a Mach-Zehnder interferometer (MZI) with two unknown phases in its upper and lower arms, as shown in Fig.~\ref{fig:1}, and propose a scheme for simultaneously estimating the two phases, achieving Heisenberg-scaling precision in both parameters. In particular, we inject squeezed and coherent states into the input ports and employ two homodyne detectors at the output ports of the MZI. We demonstrate that our scheme does not impose any constraint on the values of the parameters and the sensitivity reaches Heisenberg scaling regardless of the parameter values. Also, our scheme does not require parameter-dependent adaptation of the optical network. Not only does our model enable reaching the Heisenberg-scaling multiparameter Cram\'{e}r-Rao bound (CRB) for both unknown parameters, but it is also experimentally realizable due to the use of feasible Gaussian states and homodyne measurements.

This paper is arranged as follows. In Sec.~\ref{SectionII} we introduce our sensing scheme based on an MZI with two unknown phases. In Sec.~\ref{SectionFI} we give an analysis of the Fisher information and discuss the possibility of achieving the Heisenberg scaling. In Sec.~\ref{SectionIII} we estimate an arbitrary linear combination of two parameters by only processing the homodyne signal. In Sec.~\ref{SectionIV} we show the Heisenberg scaling in both parameters simultaneously by calculating the Fisher information matrix. We conclude by discussing our results in Sec.~\ref{SectionV}.
\section{Sensing scheme: Mach-Zehnder interferometer}
\label{SectionII}
\begin{figure}[t]
    \centering
    \includegraphics[width=1\linewidth]{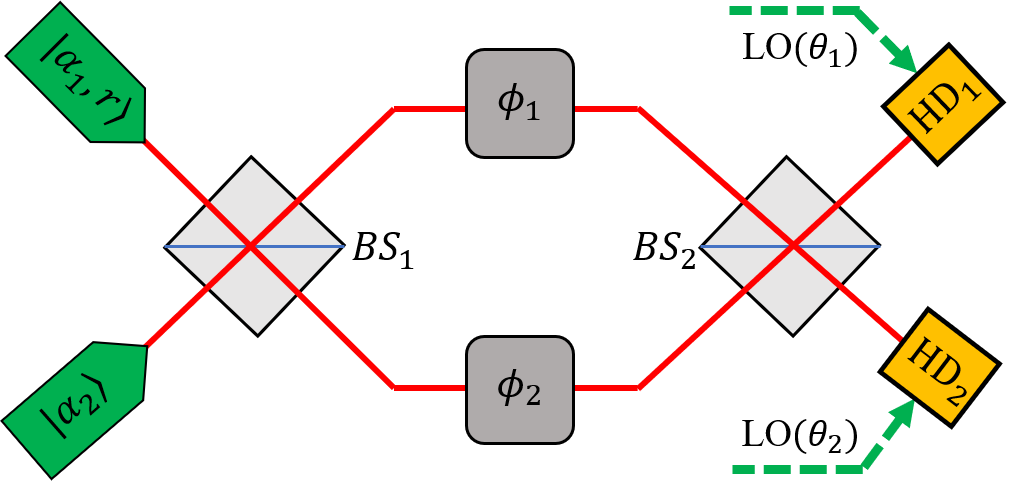}
    \caption{Schematic of an MZI for the simultaneous measurement of the sum and differential phases. The input probe is prepared with a displaced squeezed state and a coherent state of light $\ket{\psi}_{\mathrm{in}}=\ket{\alpha_1,r}\otimes\ket{\alpha_2}$ as defined in Eq.~\ref{eq:INPUTstate}. The two output modes are measured using two homodyne
detectors (HDs) to perform a joint measurement with local oscillator (LO) phases $\theta_1$ and $\theta_2$.}
    \label{fig:1}
\end{figure}

The MZI is a two-channel linear passive optical device consisting of two beam splitters (BS) and phase shifts in its arms. We consider the beam splitters to be 50:50 and the phase shift $\phi_1$ ($\phi_2$) in the upper (lower) arm of our MZI arbitrary, as depicted in \figurename~\ref{fig:1}. The action of a 50:50 BS and of two unknown phase shifts on the two optical modes can be represented respectively by the unitary matrices
\begin{equation}
\begin{split}
&U_{\mathrm{BS}}=\exp\left(\i\frac{\pi}{4}\sigma_y\right)=\frac{1}{\sqrt{2}}\begin{pmatrix}
1&1\\-1&1
\end{pmatrix},\\
&U_{\mathrm{PH}}(\phi_1,\phi_2)=\begin{pmatrix}
e^{\i\phi_1} & 0\\
0& e^{\i\phi_2}\\
\end{pmatrix} ,
\end{split}
\label{eq:UnitaryMatrix}
\end{equation}
where $\sigma_y$ is the second Pauli matrix. Hence the overall network depicted in \figurename~\ref{fig:1} is described by
\begin{equation}
U=U_{\mathrm{BS}}^{\dagger}U_{\mathrm{PH}}U_{\mathrm{BS}} .
\label{eq:MZIUnitary}
\end{equation}
The entries of this unitary matrix can be written as $U{_{ij}}=\sqrt{p_{ij}}e^{\i \gamma_{ij}}$, where $p_{ij}=|U_{ij}|^2$ is the probability that a photon in the input channel $j$ propagates into the output channel $i$ of the interferometer, and $\gamma_{ij}=\mathrm{Arg}[U_{ij}]$ is the phase that the same photon would acquire.
%Since $U$ is a $2\times 2$ unitary matrix, 
It is easy to check that $p_{11}=p_{22}=p_1$ and $p_{12}=p_{21}=p_2$.

We now propose a scheme that can achieve the Heisenberg scaling in the simultaneous estimation of two unknown parameters, namely the sum $\phi_s$ and the difference $\phi_d$ of the phases in the MZI. As shown in \figurename~\ref{fig:1}, firstly we inject %a displaced squeezed state and a coherent state of light as input probe. In particular, our scheme employs the input state represented as
as input probe the state
\begin{equation}
\ket{\psi}_{\mathrm{in}}=\ket{\alpha_1,r}\otimes\ket{\alpha_2}.
\label{eq:INPUTstate}
\end{equation}
This state comprises a displaced squeezed state $\ket{\alpha_1,r}$ with an average number of photons $N_{c1}+N_s=|\alpha_1|^2+\sinh^2{r}$, and a coherent state $\ket{\alpha_2}$ with an average number of photons $N_{c2}=|\alpha_2|^2$, allocated respectively to the first and second input ports of the MZI. Here $r$ represents the real squeezing parameter,  and $\alpha_1$ and $\alpha_2$ denote the real amplitudes of the coherent light contributions to the input states.

Then we perform homodyne detections at both output ports to measure the field quadratures $\hat{x}_{i,\theta_i}$, with $i=1,2$, where $\theta_i$ is the local oscillator reference phase of $i$-th homodyne detector, from which we infer the value of the unknown parameters $\bm{\phi}=(\phi_s,\phi_d)$.
The joint probability distribution $p_{\bm{\phi}}(\vec{x})$ associated with the two homodyne detections at the output ports is Gaussian and reads
\begin{equation}
p_{\bm{\phi}}(\vec{x})=\frac{1}{2\pi\sqrt{\mathrm{Det}[\Sigma]}}\mathrm{exp}\Bigg[-\frac{(\vec{x}-\vec{\mu})^T\Sigma^{-1}(\vec{x}-\vec{\mu})}{2}\Bigg],
\label{eq:ProbDist}
\end{equation}
where $\Sigma$ is the second moment and $\vec{\mu}$ is the first moment that encode the unknown parameters $\phi_s$ and $\phi_d$ through the interferometric evolution of the input state. The detailed expressions of the first and second-order moments of the output state are presented in Appendix~\hyperref[firstappendix]{A}. Since we are measuring both the output ports of the MZI with the probes in both input ports, all the elements of the unitary matrix in Eq.~\eqref{eq:MZIUnitary} will be relevant. The mean $\vec{\mu}$ and the elements of the covariance matrix $\Sigma$ can be written, relatively to the phases of the two local oscillators, as %(see Appendix~\hyperref[firstappendix]{A})
\begin{equation}
\vec{\mu}=\sqrt{2}\begin{pmatrix}
\alpha_1\sqrt{p_1}\cos(\gamma_{11}-\theta_1)+\alpha_2\sqrt{p_2}\cos(\gamma_{12}-\theta_1)\\
\alpha_1\sqrt{p_2}\cos(\gamma_{21}-\theta_2)+\alpha_2\sqrt{p_1}\cos(\gamma_{22}-\theta_2)
\end{pmatrix}
\label{eq:Disp}
\end{equation}
and 
\begin{equation}
\begin{split}
&\Sigma_{11}=\frac{1}{2}+p_1[\sinh^{2}r+\cos2(\gamma_{11}-\theta_1)\sinh{r}\cosh{r}]\\
&\Sigma_{22}=\frac{1}{2}+p_2[\sinh^{2}r-\cos2(\gamma_{21}-\theta_2)\sinh{r}\cosh{r}]\\
&\Sigma_{12}=\Sigma_{21}=\sqrt{p_1p_2}[\cos(\gamma_{11}-\gamma_{21}-\theta_1+\theta_2)\sinh^{2}r\\&          +\cos(\gamma_{11}+\gamma_{21}-\theta_1-\theta_2)\sinh(r)\cosh(r)].
\end{split}
\label{eq:CovElement}
\end{equation}
We note that the measurable homodyne signal depends on the internal interferometer phases $\phi_s$ and $\phi_d$ relative to the LO phases. Therefore, although the interferometer itself remains a two-mode device, the effective estimation scheme includes external reference phases through the LOs. Furthermore, the LO is treated as a classical reference with effectively infinite amplitude, which does not constrain the Fisher information, as shown in the next sections.

\section{Fisher information analysis}
\label{SectionFI}

The Fisher information quantifies the amount of information about the unknown parameters to be estimated which can be retrieved with a given measurement scheme. Here, we calculate the Fisher information matrix (FIM) associated with the estimation of both phase shifts $\phi_s$ and $\phi_d$ by performing homodyne detections at both output ports of the MZI. The FIM gives a lower bound on the estimation uncertainty, known as the Cram\'{e}r-Rao bound, given by
\begin{equation}
\mathrm{Cov}[\tilde{\bm{\phi}}]\geq \frac{1}{\nu}F^{-1}[\bm{\phi}],
\label{eq:CRB}
\end{equation}
where $\nu$ is the number of iterations of the measurement, $F$ is the positive semi-definite Fisher information matrix, and $\mathrm{Cov}[\tilde{\bm{\phi}}]$ denotes the covariance matrix associated with the estimators $\tilde{\bm{\phi}}=(\tilde{\phi}_s, \tilde{\phi}_d)$ of the sum and difference of the two phases. By employing the probability distribution~\eqref{eq:ProbDist}, the entries of the FIM associated with our scheme have been shown in Appendix~\hyperref[Appendix B]{B} and read
\begin{equation}
F_{mn}=\underbrace{\partial_{\phi_m}\vec{\mu}^{T}\Sigma^{-1}\partial_{\phi_n}\vec{\mu}}_{\text{$F^\mathcal{S}_{m n}$}}+\underbrace{\frac{1}{2}\mathrm{Tr}\Big[\Sigma^{-1}(\partial_{\phi_m}\Sigma)\Sigma^{-1}(\partial_{\phi_n}\Sigma)\Big]}_{\text{$F^\mathcal{N}_{m n}$}},
\label{eq:FIMEqn}
\end{equation}
for $m,n\in\{s,d\}$, where $\mathrm{Tr}[\cdot]$ denotes the trace operation. Since the MZI employs a displaced squeezed state and a coherent state, the FIM in Eq.~\eqref{eq:FIMEqn} is the sum of two contributions $F^\mathcal{S}$ and $F^\mathcal{N}$, associated respectively with the variations of the \emph{signal} and of the \emph{noise} of the outcome of homodyne measurements. 

%Furthermore, the entries of $F^\mathcal{N}$ can be  rewritten as (see Appendix~\hyperref[Appendix B]{B}) \begin{equation} F^\mathcal{N}_{mn}=\frac{(\partial_{\phi_m}\mathrm{Det}[\Sigma])(\partial_{\phi_n}\mathrm{Det}[\Sigma])}{2\mathrm{Det}[\Sigma]^2}-\frac{\mathrm{Tr}\Big[(\partial_{\phi_m}C)(\partial_{\phi_n}\Sigma)\Big]}{2\mathrm{Det}[\Sigma]},\label{eq:FIMsqueezed} \end{equation} in terms of the cofactor matrix $C=\mathrm{Det}[\Sigma]\Sigma^{-1}$ and the determinant Det[$\Sigma$] of the covariance matrix. 
%n order to evaluate the asymptotic behaviour of the FIM, we first notice that the the elements of the covariance and cofactor matrices scale as $N$, namely $C_{ij}=O(N_s)$, $\Sigma_{ij}=O(N_s)$ and Det[$\Sigma$]$=O(N_s)$. The same asymptotic behaviour are also shown by their respective derivatives, while $\partial_{\phi_i}\vec{\mu}$ is of order $O(\sqrt{N_{c}})$, since neither of $p_i$ and $\gamma_{ij}$ depends on $N_s$ or $N_{c}$. By inserting these quantities in Eq.~\eqref{eq:FIMEqn} it is clear that none of the elements of the FIM scale quadratically in the photon number (see Appendix~\hyperref[Appendix B]{B}). Therefore, i

Let us denote with $N_c=N_{c1}+N_{c2}$ the total average number of photons in the displacement of the probes, and with $N=N_c+N_s$ the total average number of photons injected into the MZI. In the next  sections, we will show, by assuming $N_{c,s}= O(N)$, that it is possible to achieve Heisenberg-scaling precision $O(1/N)$ in the estimation of the parameters $\phi_s$ and $\phi_d$ without any adaptation of the optical network by simply tuning experimentally the local oscillators in the homodyne measurements according to the conditions
\begin{equation}
\theta_i=\gamma_{i1}+\frac{\pi}{2}+\frac{k_i}{N_s},\quad  i=1,2 .
\label{eq:ConditionLO}
\end{equation}
Here, $\gamma_{i1}+\pi/2$ are the phases of quadrature fields $\hat{x}_{i,\gamma_{1i}+\pi/2}$ with minimum variance, differing from the local oscillator phases by a term $k_i/N_s$, with $k_i$ being a real constant independent of $N_s$.

In the next section, we will show that the information $F^\mathcal{S}$ in Eq.~\eqref{eq:FIMEqn} associated with the main homodyne signal is enough to achieve Heisenberg-scaling precision in the estimation of a linear combination of the two unknown phase parameters with arbitrary positive weights. By taking advantage also of the information $F^\mathcal{N}$ associated with the fluctuations of the signal is further possible to estimate simultaneously both parameter with Heisenberg scaling precision, as we will show in Sec.~\ref{SectionIV}.

\section{Estimation of a linear combination of phase parameters}
\label{SectionIII}
Here we analyse the first contribution to the FIM~\eqref{eq:FIMEqn} which is the information extracted from the variation of the displacement $\vec{\mu}$ with respect to the parameters $\phi_s$ and $\phi_d$.
%is given by 
%\begin{equation}
%F^\mathcal{S}=\begin{pmatrix}
%F^\mathcal{S}_{ss}&F^\mathcal{S}_{sd}\\F^\mathcal{S}_{ds}&F^\mathcal{S}_{dd}
%\end{pmatrix}
%\end{equation}
In the condition~\eqref{eq:ConditionLO} for $k_1=k_2=k$ and for a large number of photons it asymptotically  reads (see Appendix~\hyperref[Appendix C]{C})
\begin{equation}
F^\mathcal{S}=\frac{4N_s N_c}{(16k^2+1)}\begin{pmatrix}
\beta&\sqrt{\beta(1-\beta)}\\ \sqrt{\beta(1-\beta)}& 1-\beta
\end{pmatrix},
\label{eq:FIMdisp}
\end{equation}
where $\beta=N_{c1}/N_c$ is the ratio of coherent photons in the first channel. $F^\mathcal{S}$ comes out to be a singular matrix, proportional to $(\sqrt{\beta}, \sqrt{1-\beta})^T (\sqrt{\beta}, \sqrt{1-\beta})$, which means that only one parameter can be estimated with Heisenberg-scaling precision.  The bound in Eq.~\eqref{eq:CRB} is not well defined due to the non-invertible FIM~\cite{stoica2001parameter} (see Appendix~\hyperref[Appendix:Psedoinverse]{D}). However, in the condition~\eqref{eq:ConditionLO} we can estimate the linear combination $\Phi= \sqrt{\beta}\phi_s +\sqrt{1-\beta} \phi_d$ of the parameters with Heisenberg-scaling precision 
 \begin{equation}
 (\Delta\Phi)^2\geq \frac{1}{\nu}\frac{16k^2+1}{4N_s N_c}.
 \label{eq:CRBDispl}
\end{equation}  
It is worth noticing that the linear combination $\Phi$ to be estimated can be arbitrarily selected by varying the intensities of the displacement in the input ports. If we set $\beta=0$, i.e.,\ only a squeezed vacuum state and a coherent state are injected as input, one can estimate the parameter $\phi_d$ with Heisenberg-scaling sensitivity. This configuration is a very important application of MZI and is often considered for gravitational wave detection~\cite{aasi2013enhanced}, and has been explicitly studied in several works~\cite{PhysRevLett.100.073601, PhysRevLett.111.173601}. On the other hand, if we set $\beta=1$, i.e.,\ we keep the second input port vacuum and a displaced squeezed state is injected in the first port, we can estimate the parameter $\phi_s$ with the Heisenberg-scaling precision, whereas the sensitivity in the parameter $\phi_d$ cannot exceed the shot-noise scaling~\cite{takeoka2017fundamental}.

\section{Two-parameter estimation}\label{SectionIV}

We now demonstrate that the estimation of \emph{both} MZI phase parameters simultaneously achieves the Heisenberg scaling. This goal is obtained by using also the information given by the second term $F^\mathcal{N}$ in the FIM~\eqref{eq:FIMEqn} which is associated with the variation of the noise. This matrix gives only the information on the parameters obtained from the variation of the squeezing of the probe, that is independent of the displacement. 
Assuming condition~\eqref{eq:ConditionLO} for $k_1=k_2=k$ and for a large number of photons, we have (see Appendix~\hyperref[Appendix C2]{C~2})
\begin{equation}
F^{\mathcal{N}}=\begin{pmatrix}
\frac{128N_{s}^2k^2}{(16k^2+1)^2}&0\\ 0&0
\end{pmatrix},
\label{eq:FIMsqueez}
\end{equation}
neglecting all orders smaller than $O(N_s^2)$ which do not affect the Heisenberg scaling. Therefore, the sensitivity in the estimation of $\phi_s$ in Eq.~\eqref{eq:CRB}, reads
\begin{equation}
(\Delta\phi_s)^2\geq \frac{1}{\nu}\frac{(16k^2+1)^2}{128N_s^2k^2} ,
   \label{eq:squeezCRBPhi_s}
\end{equation}
allowing the estimation of the parameter $\phi_s$ with Heisenberg scaling, whereas the sensitivity in the parameter $\phi_d$ cannot be better than the SNL.

It is evident that one can estimate only one parameter at a time with Heisenberg scaling if the information of the parameters is retrieved only from the signal or only from the fluctuation.
In order to truly perform multi-parameter estimation, we need the complete information, retrieved from both the variations of the signal and of the signal fluctuations with respect to the parameters, expressed by the whole FIM $F=F^\mathcal{S}+F^\mathcal{N}$ from Eqs.~\eqref{eq:FIMdisp} and~\eqref{eq:FIMsqueez}, and reads

\begin{equation}
F=\frac{4N_sN_c}{16 k^2+1}\left(
\begin{array}{cc}
\beta+\frac{32N_sk^2}{N_c(16k^2+1)} & \sqrt{\beta (1-\beta)} \\
  \sqrt{\beta(1-\beta)} & (1-\beta ) \\
\end{array}
\right).
\label{eq:FullFIM}
\end{equation}
The above FIM gives Heisenberg-scaling precision in the simultaneous estimation of $\phi_s$ and $\phi_d$. In the Appendix~\ref{Appendix C3}, we show that the total sensitivity $(\Delta\phi_s)^2+(\Delta\phi_d)^2=\mathrm{Tr}[F^{-1}]$ is maximised for $\beta=0$, which makes the FIM in Eq.~\eqref{eq:FullFIM} a diagonal matrix, given by
\begin{equation}
F=\begin{pmatrix}
\frac{128N_s^2k^2}{(16k^2+1)^2} &0 \\
0 &\frac{4 N_sN_c}{16 k^2+1} \\
\end{pmatrix},
\label{eq:DiagMatrix}
\end{equation}
where the upper diagonal element stems only from $F^\mathcal{N}$ in Eq.~\eqref{eq:FIMsqueez} and the lower diagonal element stems only from $F^\mathcal{S}$ in Eq.~\eqref{eq:FIMdisp} for $\beta=0$. The sensitivities in the simultaneous estimation of $\phi_s$ and $\phi_d$ given by the CRB in Eq.~\eqref{eq:CRB}, read
\begin{align}
(\Delta\phi_s)^2&\geq
(\Delta\phi_s^{CRB})^2= \frac{1}{\nu}\frac{(16k^2+1)^2}{128N_s^2k^2} ,
   \label{eq:CRBPhi_s}
\\
(\Delta\phi_d)^2&\geq(\Delta\phi_d^{CRB})^2=\frac{1}{\nu}
\frac{(16k^2+1)}{4N_cN_s}.
\label{eq:CRBPhi_d}
\end{align}
We notice that the CRB in Eq.~\eqref{eq:squeezCRBPhi_s} asymptotically coincide with the one in Eq.~\eqref{eq:CRBPhi_s} whereas Eq.~\eqref{eq:CRBPhi_d} coincide with the bound in Eq.~\eqref{eq:CRBDispl} for $\Phi = \phi_d$ (for $\beta =0$), suggesting that in the Heisenberg-scaling estimation of $\phi_s$ and $\phi_d$, the dominant contributions for large $N$ come respectively from the Fisher information terms $F^\mathcal{N}$ and $F^\mathcal{S}$. 
\begin{figure}
    \centering
    \includegraphics[width=85mm]{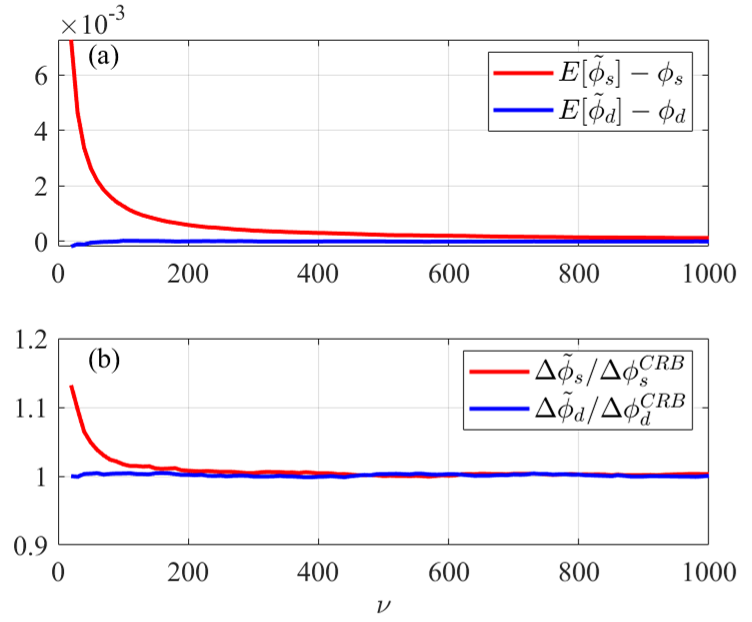}
    \caption{Estimation as a function of the number of measurements $\nu$. (a) Bias and (b) Saturation of the CRB for the estimated parameters obtained from the maximum likelihood. Both parameters $\phi_s$ (red) and $\phi_d$ (blue) in Eq.~\eqref{eq:CRBPhi_s}--\eqref{eq:CRBPhi_d} saturate the bound for a large number of repetitions of the measurement. Here, $\alpha_1=0$, $|\alpha_2|^2=10$ and $r=1.7$.}
    \label{Fig2}
\end{figure}
\begin{figure}
    \centering
   \includegraphics[width=85mm]{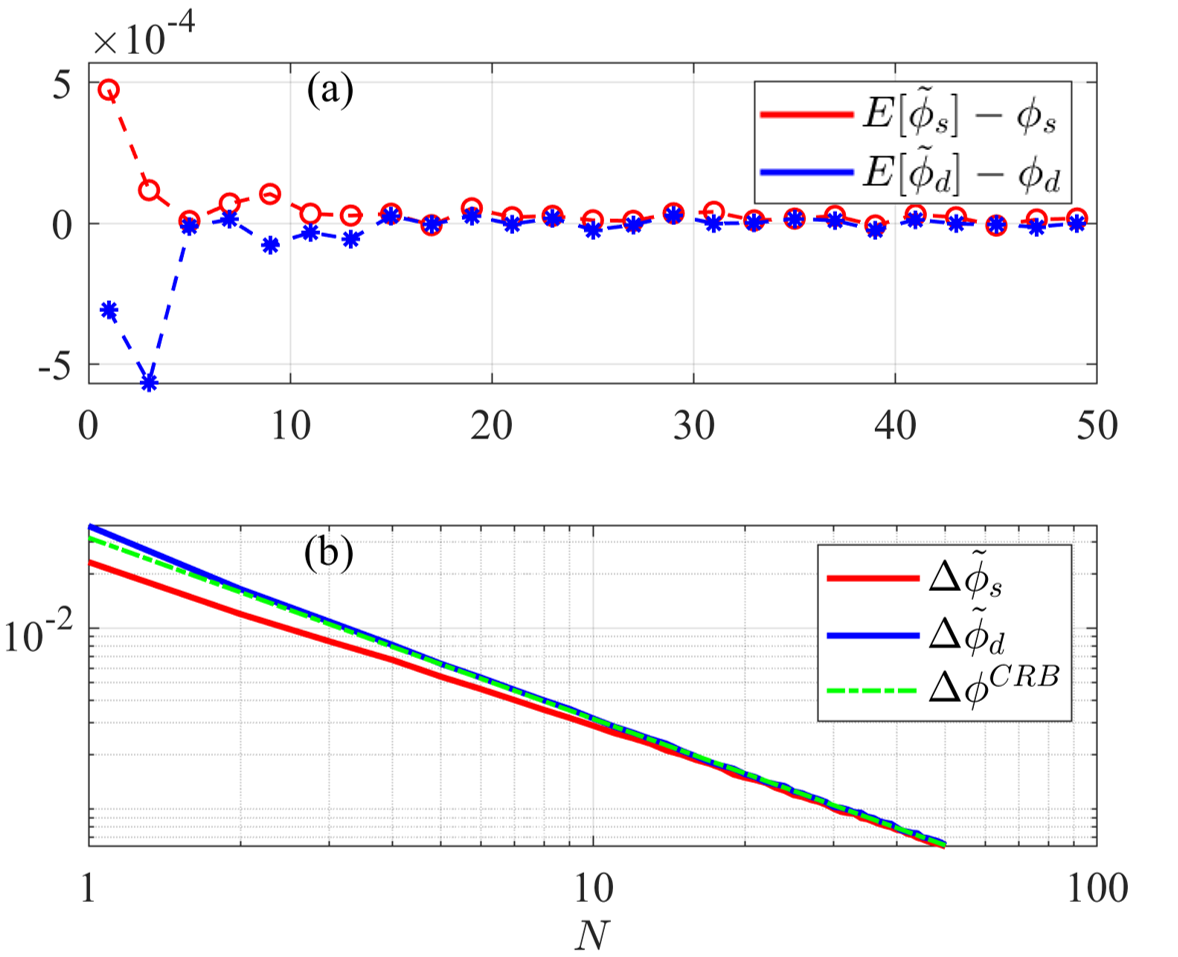}
    \caption{Estimation as a function of total number of photons $N$, with equal average number of photons in squeezed and coherent states, $|\alpha_2|^2=\sinh^2{r}=N/2$, considering $\alpha_1=0$. (a) Bias of the estimated parameters $\phi_s$ (red) and $\phi_d$ (blue). (b) Demonstration of the Heisenberg scaling $\Delta\phi_{s(d)}\sim 1/N$. The CRB saturated for very few number of photons. The uncertainty in the estimation of $\phi_s$ (red) and $\phi_d$ (blue) is plotted with the CRB (green) defined in Eq.~\eqref{eq:CRBPhi_s}--\eqref{eq:CRBPhi_d}, where $\Delta\phi^{CRB}=\Delta\phi_{s(d)}^{CRB}$ for $k=1/4$. Here, the sample size $\nu=2000$.}
    \label{Fig3}
\end{figure}
Noticeably, the FIM in Eq.~\eqref{eq:FIMEqn} in the asymptotic regime gives a Heisenberg scaling precision in the estimation of both parameters. Moreover, $\phi_s$ achieves Heisenberg scaling with $\Delta\phi_s=O( 1/N_s)$, irrespective of the intensity of the coherent state, and depends only on the squeezed photons. On the other hand, the estimation of $\phi_d$ achieves Heisenberg scaling, provided that both the average number of photons in the coherent light $N_c$ and in the squeezing $N_s$ scale with the total average number of photons $N = N_s + N_c$, and is optimal when $N_c=N_s=N/2$ (i.e., with half of the input intensity provided by the coherent light and half by the squeezed light), giving $\Delta\phi_d=O(1/N)$. We point out that the quantum correlation between the upper and lower channels introduced by the first BS has a key role in achieving the Heisenberg scaling in the simultaneous estimation of both $\phi_s$ and $\phi_d$. Indeed, without the first BS, it would be possible to achieve Heisenberg scaling only in the estimation of the phase $\phi_1$.

The Cram\'{e}r-Rao bounds in Eq.~\eqref{eq:CRBPhi_s}--\eqref{eq:CRBPhi_d} are asymptotically saturated by employing the maximum-likelihood estimator, thus achieving Heisenberg-scaling precision.  
The expressions of the maximum-likelihood estimators for this estimation scheme are derived in Appendix~\ref{Appendix E}, and their behavior is shown in Figs.~\ref{Fig2} and~\ref{Fig3} as a function of the number of probe photons and the number of experimental runs. Figs.~\ref{Fig2}(a) and~\ref{Fig3}(a) represent the bias of the estimated parameters defined as the difference between the expected value of the estimator and the true value of the parameter (i.e., $\mathrm{E}[\Tilde{\phi}_{s(d)}]-\phi_{s(d)}$, where $\mathrm{E}[\cdot]$ denotes the expectation operation). In Figs.~\ref{Fig2}(b) and~\ref{Fig3}(b) we show instead the uncertainties $\Delta\phi_{s(d)}$ of the estimators with respect to the Cram\'{e}r-Rao bounds as in Eq.~\eqref{eq:CRBPhi_s}--\eqref{eq:CRBPhi_d}. We observe that the estimators $\Tilde{\phi}_s$ and $\Tilde{\phi}_d$ remain unbiased and provide a good approximation of the Cram\'{e}r-Rao bound with Heisenberg scaling already for relatively small values of experimental iterations and of the average number of photons in the squeezed and coherent states.

\section{Conclusions}
\label{SectionV}
We have presented an experimentally feasible scheme to estimate two unknown parameters with ultimate sensitivity simultaneously in an MZI illuminated by a squeezed and coherent state of light. We showed that homodyne detection in both ports allows us to estimate both phases in the two interferometer arms with Heisenberg scaling precision independently of the values of the parameters and without the need of adapting the network. Moreover, we also showed that a linear combination of phase sum and difference parameters can be estimated with the Heisenberg scaling by harnessing only the variation in the signal of the measurements.  Additionally, we demonstrated that a maximum likelihood estimator saturates the Cram\'{e}r-Rao bound already by employing only a number of experimental iterations of the order of 100.  Harnessing the quantum advantage of simultaneous Heisenberg scaling estimation of both parameters opens a new frontier in MZI-based quantum metrology and can pave the way to future schemes for multi-parameter estimation in more general optical networks for quantum technology applications with feasible experimental resources.

\acknowledgments

This work was partially supported by Xairos System Inc. VT also acknowledges partial support from the Air Force Office of Scientific Research under award number FA8655-23-17046.
PF was partially supported by Istituto Nazionale di Fisica Nucleare (INFN) through the project ``QUANTUM'', by the Italian National Group of Mathematical Physics (GNFM-INdAM), and by the Italian funding within the ``Budget MUR - Dipartimenti di Eccellenza 2023--2027'' - Quantum Sensing and Modelling for One-Health (QuaSiModO). D.T. acknowledges the Italian Space Agency (ASI, Agenzia Spaziale Italiana) through the project Subdiffraction Quantum Imaging (SQI) n. 2023-13-HH.0.

\newpage
\onecolumngrid
\appendix
\numberwithin{equation}{section}
\renewcommand{\theequation}{A\arabic{equation}}       
  \setcounter{equation}{0} 
\section{Joint probability distribution: Two homodyne detection}
\label{firstappendix}

In this Appendix, we will show the outcomes of homodyne detections performed at the output of an MZI with Gaussian states as input. An $m$-mode bosonic  system is usually described in terms of a vector of quadrature operators $\hat{R}$ =
$(\hat{q}_1,\hat{q}_2,...,\hat{q}_m,\hat{p}_1,\hat{p}_2,...,\hat{p}_m)$, which satisfy the canonical commutation relation
\begin{equation}
[\hat{R_k},\hat{R_l}]=\i\Omega_{kl},
\end{equation}
where $\Omega$ is the symplectic matrix
\begin{equation}
\Omega=\begin{pmatrix}
0&I\\-I &0
\end{pmatrix}
\end{equation}
with $I$ being an $m\times m$ identity matrix. 

The vector $\bm{d}$ of the first moments of a quantum state $\rho$ is defined as
\begin{equation}
d_k=\braket{\hat{R_k}},
\end{equation}
and the second moment also called covariance matrix $\Gamma$ is
\begin{equation}
\Gamma_{ij}=\frac{1}{2}\braket{\hat{R_i}\hat{R_j}+\hat{R_j}\hat{R_i}}-\braket{\hat{R_i}}\braket{\hat{R_i}}.
\end{equation}
Here $\braket{\hat{O}} \equiv$ Tr[$\rho \hat{O}$] denotes the mean of the operator $\hat{O}$ evaluated on the
state $\rho$. We note that the covariance matrix $\Gamma$ is a real, symmetric, positive definite
matrix. The uncertainty relation of canonical operators imposes a constraint on the covariance matrix, corresponding to the inequality
\begin{equation}
  \Gamma -\frac{i}{2}\Omega \geq 0 .
\end{equation}

 A Gaussian state is completely characterised by its first and second canonical moments $\bm{d}$ and $\Gamma$ respectively.
 An $m$-mode Gaussian state $\rho$ can be completely determined through its Wigner distribution function, which is Gaussian, given specifically by
\begin{equation}
    W(\bm{z})=\frac{1}{(2\pi)^2 \sqrt{\det[\Gamma]}}\exp{\left(-\frac{1}{2}(\bm{z}-\bm{d})^T\Gamma^{-1}(\bm{z}-\bm{d})\right)},
\end{equation}
with $\bm{z}\in\mathbb{R}^{2m}$.

Here in particular we discuss a 2-channel passive linear optical network described by unitary matrix $\hat{U}$. Let us consider a displaced squeezed state in the first channel and a coherent state in the second input channel given by $\ket{\alpha_1,r}\otimes \ket{\alpha_2}$. 
We assume the amplitudes $\alpha_1$ and $\alpha_2$ of the coherent state and the squeezing coefficient $r$ to be real. The covariance matrix of this product state is given by
\begin{equation}
\Gamma=\frac{1}{2}\left(
\begin{array}{cccc}
 e^{2r} & 0 &
 0 & 0 \\
 0 & 1 & 0 &
 0 \\
0 & 0 & e^{-2r} & 0 \\
 0 & 0 & 0 & 1\\
\end{array}
\right),
\label{eq:APXAsigma}
\end{equation}
or equivalently $\Gamma=\frac{1}{2}\diag(\e^{2S},\e^{-2S})$, where $S= \diag(r,0)$.
 The displacement vector is $\bm{d}=\sqrt{2} (\alpha_1,\alpha_2,0,0)$. The passivity and linearity of the network preserve the Gaussian nature of the state.
The evolution of the state after the action of a passive and linear unitary $\hat{U}$ on an initial state is described by a symplectic
rotation of the state Wigner distribution. The Wigner distribution $W(\bm{z})$ of the state
$\hat{U}\ket{\alpha_1,r}\otimes \ket{\alpha_2}$ after the interferometric evolution $\hat{U}$ is thus obtained by rotating the initial Wigner
distribution $W(\bm{z})$ %found in Eq.~\eqref{eq:MZIUnitary} 
with the orthogonal and symplectic matrix
\begin{equation}
     R= \begin{pmatrix}
\Re[U]& -\Im[U]\\
\Im[U] & \Re[U]
\end{pmatrix},
\label{EqnA8}
\end{equation}
where $U$ is the unitary matrix representation of the passive and linear interferometric evolution $\hat{U}$ that satisfies $\hat{U}^\dag\hat{a}^\dag_i\hat{U}=\sum_{j=1,2}U_{ij}\hat{a}^\dag_j$, with $\hat{a}_i$ bosonic mode of the $i$th channel and $i=1,2$.
The Wigner function after the rotation reads
\begin{equation}
    W_U(\bm{z})=W(R^T\bm{z}).
    \label{EqnA9}
\end{equation}
Since the probe in our model is Gaussian, the transformation of the state is completely
described by the rotations of the displacement
\begin{equation}
\begin{split}
    \bm{d}_U=R\bm{d}=&\begin{pmatrix}
\Re[U]& -\Im[U]\\
\Im[U] & \Re[U]
\end{pmatrix}\times\sqrt{2}\begin{pmatrix}
    \alpha_1\\ \alpha_2\\0\\0
\end{pmatrix},
\label{eq:Rotaiondisp}
\end{split}
\end{equation}
 and of the covariance matrix $\Gamma$~\eqref{eq:APXAsigma} of the initial state
\begin{equation}
  \Gamma_U=R\Gamma R^T=\begin{pmatrix}
        \Delta X^2 & \Delta XP\\
        \Delta XP^T & \Delta P^2
    \end{pmatrix}.
    \label{EqnA11}
\end{equation}
 Here $\Delta X^2$, $ \Delta XP$ and $\Delta P^2$ are the $2\times2 $ matrices
        \begin{equation}
        \begin{split}
            \Delta X^2&=\frac{1}{2}[\Re[U]e^{2S}\Re[U^\dagger]-\Im[U]e^{-2S}\Im[U^\dagger]]\\
            &=\frac{1}{2}[\Re[U \cosh(2S)U^\dagger]+\Re[U 
 \sinh(-2S)U^T]],
            \end{split}
            \label{EqnA12}
        \end{equation}
         \begin{equation}
        \begin{split}
            \Delta P^2&=\frac{1}{2}[-\Im[U]e^{2S}\Im[U^\dagger]+\Re[U]e^{-2S}\Re[U^\dagger]]\\
            &=\frac{1}{2}[\Re[U \cosh(2S)U^\dagger]-\Re[U\sinh(-2S)U^T]],
            \end{split}
            \label{EqnA13}
        \end{equation}
         \begin{equation}
        \begin{split}
            \Delta XP&=\frac{1}{2}[-\Re[U]e^{2S}\Im[U^\dagger]-\Im[U]e^{-2S}\Re[U^\dagger]]\\
            &=\frac{1}{2}[-\Im[U \cosh(2S)U^\dagger]+\Im[U \sinh(-2S)U^T]].
            \end{split}
            \label{EqnA14}
        \end{equation}
By exploiting the relations
\begin{equation}
    \begin{split}
&\cosh(2S)_{ij}=\delta_{i1}\delta_{j1}\cosh(2r)+\delta_{i2}\delta_{j2} ,\\
&\sinh(2S)_{ij}=\delta_{i1}\delta_{j1}\sinh(2r),
\end{split}
\label{EqnA15}
\end{equation}
where $\delta_{ij}$ is the Kronecker delta, we get
\begin{equation}
    \begin{split}
&(U\cosh(2S)U^\dagger)_{ij}=U_{i1}U_{j1}^{*}\cosh(2r)+U_{i2}U_{j2}^{*},\\
&(U\sinh(2S)U^T)_{ij}=U_{i1}U_{j1}\sinh(2r).
\end{split}
\label{EqnA16}
\end{equation}

Finally, we perform balanced homodyne detections to measure the quadratures $\hat{x}_{i,\theta_i}$, where $\theta   =   (\theta_1,\theta_2)$ are the phases of the local oscillators at each output channel.
We can model this measurement by adding a further unitary rotation $U(\theta)$ = $\diag(\e^{-\i\theta_1},\e^{-\i\theta_2})$ on the
state $\hat{U}\ket{\alpha_1,r}\otimes\ket{\alpha_2}$, whose effect is to add a phase $\e^{-\i\theta_i}$ to every element of the $i$th row of $U$.
We can thus parametrize the elements of the overall unitary matrix as $\sqrt{p_{ij}} \e^{\i(\gamma_{ij}-\theta_i)}$ where $\gamma_{ij}=\arg[U_{ij}]$ for $i,j = 1,2$.

Since we are performing homodyne detection in both output channels, the probability density function $p(x|\bm{\phi})$ of the joint measurement will be a Gaussian distribution with mean vector $\vec{\mu}$ in Eq.~\eqref{eq:Disp}, given by the first 2 elements of $\bm{d}_U$ from Eq.~\eqref{eq:Rotaiondisp},
\begin{equation}
  \vec{\mu}=\sqrt{2}\begin{pmatrix}
\alpha_1\sqrt{p_1}\cos(\gamma_{11}-\theta_1)+\alpha_2\sqrt{p_2}\cos(\gamma_{12}-\theta_1)\\
\alpha_1\sqrt{p_2}\cos(\gamma_{21}-\theta_2)+\alpha_2\sqrt{p_1}\cos(\gamma_{22}-\theta_2)
\end{pmatrix},
\label{eq:mu}
\end{equation}
and a covariance matrix $\Sigma$ in Eq.~\eqref{eq:CovElement}, given by the first 2 rows and columns of the $\Gamma_{U}$ that can be calculated from Eq.~\eqref{EqnA12}
\begin{equation}
\begin{split}
&\Sigma_{11}=\frac{1}{2}+p_1\Big[\sinh^{2}r+\cos{2(\gamma_{11}-\theta_1)}\sinh{r}\cosh{r}\Big],\\
&\Sigma_{22}=\frac{1}{2}+p_2\Big[\sinh^{2}r-\cos2(\gamma_{21}-\theta_2)\sinh{r}\cosh{r}\Big],\\
&\Sigma_{12}=\Sigma_{21}=\sqrt{p_1p_2}\Big[\cos(\gamma_{11}-\gamma_{21}-\theta_1+\theta_2)\sinh^{2}r  +\cos(\gamma_{11}+\gamma_{21}-\theta_1-\theta_2)\sinh(r)\cosh(r)\Big].
\end{split}
\label{EqnA18}
\end{equation}
Finally, the expression of the determinant of the covariance matrix can be simplified using some trigonometry and the fact that $p_1+p_2=1$, due to the unitary of the network,
\begin{equation}
\mathrm{det}[\Sigma]=\frac{1}{4}+\frac{\sinh^2{r}}{2}\Big[1-p_1p_2\sin^2{(\gamma_{11}-\gamma_{21}-\theta_1+\theta_2)}\Big]+\frac{\sinh{r}\cosh{r}}{2}\Big[p_1\cos{2(\gamma_{11}-\theta_1)}+p_2\cos{2(\gamma_{21}-\theta_2)}\Big].
\label{EqnA19}
\end{equation}

\section{Fisher information matrix for Gaussian probability in Eq.~\eqref{eq:FIMEqn}}
 \label{Appendix B}
 \numberwithin{equation}{section}
\renewcommand{\theequation}{B\arabic{equation}}       
  \setcounter{equation}{0} 
For the completeness, we will derive the well known FIM associated with the estimation of parameters $\bm{\phi}=\{\phi_s,\phi_d\}$ from the joint probability distribution of two-homodyne detections
\begin{equation}
p_{\bm{\phi}}(\vec{x})=\frac{1}{2\pi\sqrt{\mathrm{Det}[\Sigma]}}\mathrm{exp}\Bigg[-\frac{(\vec{x}-\vec{\mu})^T\Sigma^{-1}(\vec{x}-\vec{\mu})}{2}\Bigg],
\label{APX:ProbDist}
\end{equation}
where we assume that both $\vec{\mu}$ and $\Sigma$ depends on the parameters $\bm{\phi}$ \cite{kay1993fundamentals}. The FIM is defined as~\cite{cramer1999mathematical}
\begin{equation}
F_{mn}=\mathbb{E}_{p_{\bm{\phi}}}\Bigg[\bigg(\partial_{\phi_m}\mathrm{log}\ p_{\bm{\phi}}(\vec{x})\bigg)\bigg(\partial_{\phi_n}\mathrm{log}\ p_{\bm{\phi}}(\vec{x})\bigg)\Bigg].
\label{APXB:FIgen}
\end{equation}
We first calculate the logarithmic derivatives of $p_{\bm{\phi}}(\vec{x})$, therefore the FIM in Eq.~\eqref{APXB:FIgen} reduces to the evaluation of the expectation values of a polynomial in $(\vec{x}-\vec{\mu})$ of order up to fourth, exploiting the standard results of Gaussian integrals:
\begin{equation}
\begin{split}
\mathbb{E}_{p_{\bm{\phi}}}\big[x_i-\mu_i\big]&=0,\\
\mathbb{E}_{p_{\bm{\phi}}}\big[(x_i-\mu_i)(x_j-\mu_j)\big]&=\Sigma_{ij},\\
\mathbb{E}_{p_{\bm{\phi}}}\big[(x_i-\mu_i)(x_j-\mu_j)(x_k-\mu_k)\big]&=0,\\
\mathbb{E}_{p_{\bm{\phi}}}\big[(x_i-\mu_i)(x_j-\mu_j)(x_k-\mu_k)(x_l-\mu_l)\big]&=\Sigma_{ij}\Sigma_{kl}+\Sigma_{ik}\Sigma_{jl}+\Sigma_{il}\Sigma_{jk},
\end{split}
\label{APXB:GausInt}
\end{equation}
where we note that only even powers of term $(\vec{x}-\vec{\mu})$ contribute to the FIM, which reads
\begin{equation}
\begin{split}
F_{mn}=\frac{1}{4}&\bigg(\partial_{\phi_m}\mathrm{log}\ \mathrm{Det}[\Sigma]\bigg)\bigg(\partial_{\phi_n}\mathrm{log}\ \mathrm{Det}[\Sigma]\bigg)+\sum_{i,j=1}^{2}\bigg((\partial_{\phi_m}\vec{\mu}^T)\Sigma^{-1}\bigg)_i\bigg((\partial_{\phi_n}\vec{\mu}^T)\Sigma^{-1}\bigg)_j\mathbb{E}_{p_{\bm{\phi}}}\big[(x_i-\mu_i)(x_j-\mu_j)\big]\\
 &+\frac{1}{4}\bigg(\partial_{\phi_m}\mathrm{log}\ \mathrm{Det}[\Sigma]\bigg)\sum_{i,j=1}^2(\partial_{\phi_n}\Sigma^{-1}_{ij})\mathbb{E}_{p_{\bm{\phi}}}\big[(x_i-\mu_i)(x_j-\mu_j)\big]\\&+\frac{1}{4}\bigg(\partial_{\phi_n}\mathrm{log}\ \mathrm{Det}[\Sigma]\bigg)\sum_{i,j=1}^2(\partial_{\phi_m}\Sigma^{-1}_{ij})\mathbb{E}_{p_{\bm{\phi}}}\big[(x_i-\mu_i)(x_j-\mu_j)\big]\\
&+\frac{1}{4}\sum_{i,j,k,l=1}^2(\partial_{\phi_m}\Sigma^{-1}_{ij})(\partial_{\phi_n}\Sigma^{-1}_{kl})\mathbb{E}_{p_{\bm{\phi}}}\big[(x_i-\mu_i)(x_j-\mu_j)(x_k-\mu_k)(x_l-\mu_l)\big].
\end{split}
\label{APXB:FIexpand}
\end{equation}
We now substitute the expressions in Eq.~\eqref{APXB:GausInt} into Eq.~\eqref{APXB:FIexpand}, and using  Jacobi's formula for the derivative of the determinant of square matrices given by
\begin{equation}
\frac{\partial_{\phi_i}\mathrm{Det}[\Sigma]}{\mathrm{Det}[\Sigma]}=\mathrm{Tr}\Big[\Sigma^{-1}\partial_{\phi_i}\Sigma\Big]\equiv-\mathrm{Tr}\Big[\partial_{\phi_i}\Sigma^{-1}\Sigma\Big],
\label{APXB:Jacobi}
\end{equation}
we see that most of the terms in Eq.~\eqref{APXB:FIexpand} cancel out and the simplified form of FIM finally reads
\begin{equation}
F_{mn}=\underbrace{\partial_{\phi_m}\vec{\mu}^{T}\Sigma^{-1}\partial_{\phi_n}\vec{\mu}}_{\text{$F^\mathcal{S}_{m n}$}}+\underbrace{\frac{1}{2}\Tr\Big[\Sigma^{-1}(\partial_{\phi_m}\Sigma)\Sigma^{-1}(\partial_{\phi_n}\Sigma)\Big]}_{\text{$F^\mathcal{N}_{m n}$}},\qquad m,n\in\{s,d\}.
\label{APXB:FIMmain}
\end{equation}
The FIM in Eq.~\eqref{APXB:FIMmain} is given by the contribution of two terms, $F^\mathcal{S}$ and $F^\mathcal{N}$, the former pertaining the variation of the signal $\vec{\mu}$ and the latter the variation of the noise $\Sigma$ of the measurement outcome with respect to $\phi_s=\phi_1+\phi_2$ and $\phi_d=\phi_1-\phi_2$.

\section{Heisenberg-Scaling sensitivity} \label{Appendix C}
\numberwithin{equation}{section}
\renewcommand{\theequation}{C\arabic{equation}}       
  \setcounter{equation}{0} 
In this appendix, we will calculate the asymptotic behavior of the FIM and show that to reach Heisenberg scaling the conditions we need are the ones shown in Eq.~\eqref{eq:ConditionLO}. We first simplify the FIM $F$ by expressing the inverse of the covariance matrix $\Sigma^{-1}$ in terms of its cofactor matrix, i.e., $C=\det[\Sigma]\Sigma^{-1}$ and exploiting the Jacobi formula in Eq.~\eqref{APXB:Jacobi}, the FIM $F^\mathcal{N}$ reads
\begin{equation}
\begin{split}
F^\mathcal{N}_{mn}&=\frac{1}{2}\Tr\Big[\Sigma^{-1}(\partial_{\phi_m}\Sigma)\Sigma^{-1}(\partial_{\phi_n}\Sigma)\Big]\\
&=-\frac{1}{2}\Tr\Big[(\partial_{\phi_m}\Sigma^{-1})(\partial_{\phi_n}\Sigma)\Big]\\
&=\frac{(\partial_{\phi_m}\mathrm{det}[\Sigma])}{2\mathrm{det}[\Sigma]^2}\mathrm{Tr}\Big[C(\partial_{\phi_n}\Sigma)\Big]-\frac{1}{2\mathrm{det}[\Sigma]}\mathrm{Tr}\Big[(\partial_{\phi_m}C)(\partial_{\phi_n}\Sigma)\Big]\\
&=\frac{(\partial_{\phi_m}\mathrm{det}[\Sigma])(\partial_{\phi_n}\mathrm{det}[\Sigma])}{2\mathrm{det}[\Sigma]^2}-\frac{1}{2\mathrm{det}[\Sigma]}\mathrm{Tr}\Big[(\partial_{\phi_m}C)(\partial_{\phi_n}\Sigma)\Big].
\end{split}
\label{APXC:FNexpand}
\end{equation}
Therefore, the complete FIM reads
\begin{equation}
F_{mn}=\frac{1}{\mathrm{det}[\Sigma]}\partial_{\phi_m}\vec{\mu}^{T}C\partial_{\phi_n}\vec{\mu}+\frac{(\partial_{\phi_m}\mathrm{det}[\Sigma])(\partial_{\phi_n}\mathrm{det}[\Sigma])}{2\mathrm{det}[\Sigma]^2}-\frac{1}{2\mathrm{det}[\Sigma]}\mathrm{Tr}\Big[(\partial_{\phi_m}C)(\partial_{\phi_n}\Sigma)\Big].
\label{eq:APXBfim}
\end{equation}
From the explicit form of the covariance matrix $\Sigma$ in Eq.~\eqref{EqnA18} and the cofactor matrix $C$ with matrix elements $C_{11}=\Sigma_{22}, C_{22}=\Sigma_{11}$ and $C_{12}=-\Sigma_{21}$, it can be seen that they are at most of the order of $N_s$, similarly, their derivatives vary with $N_s$ as the probability amplitudes $p_i$, phases $\gamma_{ij}$ and their respective derivatives do not depend on $N_s$. 
The derivative of the mean vector $\vec{\mu}$ in Eq.~\eqref{eq:mu}, which appears in the first term of FIM in Eq.~\eqref{eq:APXBfim}, is written as
\begin{equation}
\begin{aligned}
\partial_{\phi_i}\vec{\mu} &=\sqrt{2}
\left(\begin{matrix}
\alpha_1\big(\partial_{\phi_i}\sqrt{p_1}\cos{(\gamma_{11}-\theta_1)}-\sqrt{p_1}\sin{(\gamma_{11}-\theta_1)}\partial{\phi_i}\gamma_{11}\big)\\
\alpha_1\big(\partial_{\phi_i}\sqrt{p_2}\cos{(\gamma_{21}-\theta_2)}-\sqrt{p_2}\sin{(\gamma_{21}-\theta_2)}\partial{\phi_i}\gamma_{21}\big)
\end{matrix}\right.\\
&\qquad\qquad
\left.\begin{matrix}
{}+\alpha_2\big(\partial_{\phi_i}\sqrt{p_2}\cos{(\gamma_{12}-\theta_1)}-\sqrt{p_2}\sin{(\gamma_{12}-\theta_1)}\partial{\phi_i}\gamma_{12}\big)\\
{}+\alpha_2\big(\partial_{\phi_i}\sqrt{p_1}\cos{(\gamma_{22}-\theta_2)}-\sqrt{p_1}\sin{(\gamma_{22}-\theta_2)}\partial{\phi_i}\gamma_{22}\big)
\end{matrix}\right)
\end{aligned}
\label{eq:DerivtiveMu}
\end{equation}
and scales at most with $\sqrt{N_{c}}$, where $N_c=N_{c1}+N_{c2}$ and $N_{cj}=|\alpha_{j}|^2$, $(j=1,2)$. 
Therefore, in order to achieve the Heisenberg scaling in the CRB, we must determine the asymptotic behavior of the determinant $\det[\Sigma]$ in Eq.~\eqref{EqnA19} appearing in all the terms in Eq.~\eqref{eq:APXBfim} and find under what conditions it cannot be of order higher than $N_s^0$.
The determinant $\det[\Sigma]$ asymptotically reads
\begin{equation}
\mathrm{det}[\Sigma]=N_sB_1+B_2+\frac{B_3}{N_s}+O\left(\frac{1}{N_s^2}\right),
\label{EqnC2}
\end{equation} 
where the factors $B_i$, $i=1,2,3$ can be simplified using some trigonometry and exploiting the fact that $p_1+p_2=1$, yielding
\begin{equation}
\begin{split}
&B_1=\frac{1}{2}\Big[\big(p_1\cos^2{(\gamma_{11}-\theta_1)}+p_2\cos^2{(\gamma_{21}-\theta_2)}\big)^2+\frac{1}{4}\big(p_1\sin{2(\gamma_{11}-\theta_1)}+p_2\sin{2(\gamma_{21}-\theta_2)}\big)^2\Big],\\
&B_2=\frac{1}{4}\Big[1+p_1\cos{2(\gamma_{11}-\theta_1)}+p_2\cos{2(\gamma_{21}-\theta_2)}\Big],\\
&B_3=\frac{1}{16}\Big[p_1\cos{2(\gamma_{11}-\theta_1)}+p_2\cos{2(\gamma_{21}-\theta_2)}\Big].\end{split}
\label{EqnC3}
\end{equation}
Now, to prevent $\det[\Sigma]$ to scale with $N_s$, the factor $B_1$ must be equal or tend to zero. 
This happens when we choose the local oscillator phases $\theta_i$ such that $\gamma_{i1}-\theta_{i}=\pi/2+n\pi$ with $n\in\mathbb{Z}$ and $i=1,2$. 
In particular, $\det[\Sigma]$ is of order $N_s^0$ or lower if $\theta_{i}=\gamma_{i1}+\pi/2+k_{i}/N_s$, with $i=1,2$. This is the condition found in Eq.~\eqref{eq:ConditionLO} in the main body. 
Under these conditions on the local oscillators, the factors $B_1$ and $B_2$ scale with $N_s^{-2}$, while the factor $B_3$ scales with $N_s^0$, thus assuring that $\det[\Sigma]$ becomes of order of $N_s^{-1}$.
Indeed, the expression for $\det[\Sigma]$ in Eq.~\eqref{EqnC2} after imposing $\theta_{i}=\gamma_{i1}+\pi/2+k_{i}/N_s$ can be obtained from Eq.~\eqref{EqnA19}, and reads
\begin{equation}
\mathrm{det}[\Sigma]\sim\frac{1}{N_s}\Big((p_1k_1+p_2k_2)^2+\frac{1}{16}\Big)=\frac{\rho}{N_s}
\label{EqnC4}
\end{equation}
However, the condition on the local oscillators $\theta_{i}=\gamma_{i1}+\pi/2+k_{i}/N_s$ also affects the scaling of the terms in the numerators of Eq.~\eqref{eq:APXBfim}.
Indeed, under this condition, we evaluate the asymptotic expressions
\begin{equation}
\partial_{\phi_i}\mathrm{det}[\Sigma]\sim 2\big[(p_1k_1+p_2k_2)(p_1\partial_{\phi_i}\gamma_{11}+p_2\partial_{\phi_i}\gamma_{21})\big],
\label{EqnC5}
\end{equation}
\begin{equation}
\Sigma_{ij}=\frac{\delta_{ij}-\sqrt{p_ip_j}}{2}+O(N_s^{-1}),
\label{EqnC6}
\end{equation}
\begin{equation}
C_{ij}=\frac{\sqrt{p_ip_j}}{2}+O(N_s^{-1}),
\label{EqnC7}
\end{equation}
and calculating the derivatives of $\Sigma$ and then imposing the condition in Eq.~\eqref{eq:ConditionLO}, the elements of $\partial_{\phi_i}\Sigma$ asymptotically reads
\begin{equation}
\begin{split}
&\partial_{\phi_i}\Sigma_{11}=-\frac{1}{2}(\partial_{\phi_i}p_1)+4k_1p_1(\partial_{\phi_i}\gamma_{11})+O(N_s^{-1}),\\
&\partial_{\phi_i}\Sigma_{22}=-\frac{1}{2}(\partial_{\phi_i}p_2)+4k_2p_2(\partial_{\phi_i}\gamma_{21})+O(N_s^{-1}),\\
&\partial_{\phi_i}\Sigma_{12}=\partial_{\phi_i}\Sigma_{21}=-\frac{1}{2}\partial_{\phi_i}(\sqrt{p_1p_2})+2\sqrt{p_1p_2}(k_1\partial_{\phi_i}\gamma_{21}+k_2\partial_{\phi_i}\gamma_{11})+O(N_s^{-1}).
\end{split}
\label{EqnC8}
\end{equation}
Similarly, the derivatives of the cofactor matrix can be obtained. 
At last we see the asymptotic behaviour of $\partial_{\phi_i}\vec{\mu}$ from Eq.~\eqref{eq:mu} scales with $\sqrt{N_c}$.

Finally, we calculate the FIM by substituting all the asymptotic terms in the Eq.~\eqref{eq:APXBfim} for two parameters $\phi_s$ and $\phi_d$, we separate our calculation in two parts: First we calculate only the displacement part of FIM, namely $F^\mathcal{S}$, then we calculate the part of FIM which is contributed only by squeezing namely  $F^\mathcal{N}$. To do so, we first specify the unitary matrix $U$ in Eq.~\eqref{eq:MZIUnitary} for the balanced MZI given by
\begin{equation}
U=e^{i\frac{\phi_s}{2}}\begin{pmatrix}
\cos{\frac{\phi_d}{2}}&i\sin{\frac{\phi_d}{2}}\\i\sin{\frac{\phi_d}{2}}&\cos{\frac{\phi_d}{2}}
\end{pmatrix}.
\label{eq:APXCMZI}
\end{equation}
We notice that the moduli of the probability amplitudes are independent of $\phi_s$, hence $\partial_{\phi_s}p_i=0$, which can be thought as a trivial consequence of the fact that the sum of the phases $\phi_s$ behaves as a global phase.
On the other hand, we can easily evaluate the derivatives $\partial_{\phi_s}\gamma_{ij}=1/2$ and $\partial_{\phi_d}\gamma_{ij}=0$. The derivative of $\vec{\mu}$ reads
\begin{equation}
\begin{split}
&\partial_{\phi_s}\vec{\mu}=\frac{1}{\sqrt{2}}\begin{pmatrix}
\sqrt{N_{c1}}\cos{\frac{\phi_d}{2}}\cos{\frac{k_1}{N_s}}+\sqrt{N_{c2}}\sin{\frac{\phi_d}{2}}\sin{\frac{k_1}{N_s}}\\ \sqrt{N_{c1}}\sin{\frac{\phi_d}{2}}\cos{\frac{k_2}{N_s}}-\sqrt{N_{c2}}\cos{\frac{\phi_d}{2}}\sin{\frac{k_2}{N_s}}
\end{pmatrix} , \\
&\partial_{\phi_d}\vec{\mu}=\frac{1}{\sqrt{2}}\begin{pmatrix}
\sqrt{N_{c2}}\cos{\frac{\phi_d}{2}}\cos{\frac{k_1}{N_s}}+\sqrt{N_{c1}}\sin{\frac{\phi_d}{2}}\sin{\frac{k_1}{N_s}}\\ \sqrt{N_{c2}}\sin{\frac{\phi_d}{2}}\cos{\frac{k_2}{N_s}}-\sqrt{N_{c1}}\cos{\frac{\phi_d}{2}}\sin{\frac{k_2}{N_s}}
\end{pmatrix}.
\end{split}
\label{eq:DMu}
\end{equation}
In the following sections, we calculate the FIM in Eq.~\eqref{eq:FIMEqn} by substituting all the elements while holding the conditions on the local oscillators.
\subsection{Fisher Information matrix $F^\mathcal{S}$} \label{Appendix C1}
The elements of FIM $F^\mathcal{S}$ in~\eqref{APXB:FIMmain} can thus be written using Eqs.~\eqref{EqnC4}, \eqref{EqnC7}, \eqref{eq:DMu} and asymptotically reads
 
\begin{equation}
\begin{split}
 &F^\mathcal{S}_{ss}=\frac{4N_sN_{c1}}{4\cos{\phi_d} \bigg((k_1-k_2)^2 \cos{\phi_d}+2
   (k_1^2-k_2^2)\bigg)+4(k_1+k_2)^2+1},\\&
  F^\mathcal{S}_{sd}= F^\mathcal{S}_{ds}=\frac{4N_s \sqrt{N_{c1}} \sqrt{N_{c2}}}{4\cos{\phi_d} \bigg((k_1-k_2)^2 \cos{\phi_d}+2
   (k_1^2-k_2^2)\bigg)+4(k_1+k_2)^2+1},\\&
 F^\mathcal{S}_{dd}=\frac{4N_sN_{c2}}{4\cos{\phi_d} \bigg((k_1-k_2)^2 \cos{\phi_d}+2
   (k_1^2-k_2^2)\bigg)+4(k_1+k_2)^2+1},
 \label{eq:APXCfim1}
 \end{split}
 \end{equation}
which is a singular matrix in the asymptotic regime for a large average number of photons. Therefore, only one parameter can be estimated with Heisenberg-scaling precision, i.e., the parameter identified by the eigenvector of the only non-zero eigenvalue of the FIM~\cite{stoica2001parameter}. The FIM $F^\mathcal{S}$ in Eq.~\eqref{eq:APXCfim1} is further simplified by assuming a symmetric case in condition~\eqref{eq:ConditionLO} for $k_1=k_2=k$ that makes $F^\mathcal{S}$ independent of the parameter $\phi_d$, given in Eq.~\eqref{eq:FIMdisp} in the main text and reads
\begin{equation}
F^\mathcal{S}=\frac{4N_s}{16k^2+1}\begin{pmatrix}
N_{c1}&\sqrt{N_{c1}N_{c2}}\\\sqrt{N_{c1}N_{c2}}&N_{c2}
\end{pmatrix}
\label{EqnC12}
\end{equation} 
A calculation of Heisenberg-scaling CRB of $F^\mathcal{S}$ in Eq.~\eqref{EqnC12} is given in Appendix~\hyperref[Appendix:Psedoinverse]{D}.

\subsection{Fisher Information matrix $F^\mathcal{N}$} \label{Appendix C2}
Next, we calculate the FIM that is the contribution of the variation of the noise $F^\mathcal{N}$ in Eq.~\eqref{APXC:FNexpand}, from Eqs.~\eqref{EqnC4}, \eqref{EqnC5}, \eqref{EqnC7} and~\eqref{EqnC8}, the elements of $F^\mathcal{N}$ asymptotically up to the order of $O(N_s^0)$ reads
\begin{equation}
\begin{split}
&F^\mathcal{N}_{ss}=\frac{32 N_s^2 \Big[(k_1-k_2) \cos{\phi_d}+k_1+k_2\Big]^2}{\Big[4 (k_1-k_2) \cos{\phi_d}
   \bigg((k_1-k_2) \cos{\phi_d}+2(k_1+k_2)\bigg)+4(k_1+k_2)^2+1\Big]^2}+O(N_s)+O(N_s^0),\\&
F^\mathcal{N}_{sd}=F^\mathcal{N}_{ds}=\frac{2N_s (k_1-k_2)\Big[(15k_1^2+18k_1k_2+15k_2^2-1)\sin{\phi_d}+
   12 (k_1^2-k_2^2) \sin{2\phi_d}+(k_1-k_2)^2\sin{3\phi_d}\Big]}{\Big[4 (k_1-k_2) \cos{\phi_d}
   \bigg((k_1-k_2) \cos{\phi_d}+2(k_1+k_2)\bigg)+4(k_1+k_2)^2+1\Big]^2}+O(N_s^0),\\&
F^\mathcal{N}_{dd}=\frac{N_s}{1+6k_1^2+6 k_2^2+4k_1k_2+8(k_1^2-k_2^2)\cos{\phi_d}+2(k_1-k_2)^2 \cos{\phi_d}}+O(N_s^0).
\label{eq:APXCfim2}
\end{split}
\end{equation}
The Cram\'{e}r-Rao inequality implies that

\begin{equation}
\begin{split}
(\Delta\phi_s)^2&\geq \frac{1}{\nu}(F^{\mathcal{N}^{-1}})_{11}=\frac{1}{\nu}\frac{F^\mathcal{N}_{dd}}{F^\mathcal{N}_{ss}F^\mathcal{N}_{dd}-F^\mathcal{N}_{sd}F^\mathcal{N}_{ds}},\\
&=\frac{\bigg(6 k_1^2+8(k_1^2-k_2^2) \cos{\phi_d}+2(k_1-k_2)^2 \cos{2\phi_d}+4k_1k_2+6 k_2^2+1\bigg)}{32 N_s^2\big[k_1+k_2+(k_1-k_2)\cos{\phi_d}\big]^2\nu}
\end{split}
\label{eq:APXCcrbS}
\end{equation}
and
\begin{equation}
\begin{split}
(\Delta\phi_d)^2&\geq \frac{1}{\nu}(F^{\mathcal{N}^{-1}})_{22}=\frac{1}{\nu}\frac{F^\mathcal{N}_{ss}}{F^\mathcal{N}_{ss}F^\mathcal{N}_{dd}-F^\mathcal{N}_{sd}F^\mathcal{N}_{ds}},\\
&=\frac{\bigg(6 k_1^2+8(k_1^2-k_2^2) \cos{\phi_d}+2(k_1-k_2)^2 \cos{2\phi_d}+4k_1k_2+6 k_2^2+1\bigg)}{\nu N_s} .
   \end{split}
   \label{eq:APXCcrbD}
\end{equation}
Therefore, the Heisenberg scaling is achieved in $F^\mathcal{N}$ only for the parameter $\phi_s$, whereas the precision in the parameter $\phi_d$ only reaches the SNL. This can be further simplified for $k_1=k_2=k$ resulting the FIM $F^\mathcal{N}$ independent of the parameter $\phi_d$, given in Eq.~\eqref{eq:FIMsqueez} in the main text up to the order of $O(N_s^2)$, reads
\begin{equation}
F^\mathcal{N}=\begin{pmatrix}
\frac{128N_s^2k^2}{(16k^2+1)^2}&0\\0&0
\end{pmatrix}.
\label{eq:APXCbothFIM}
\end{equation}

\subsection{Total Fisher Information matrix $F^\mathcal{S}+F^\mathcal{N}$}
\label{Appendix C3}

To estimate the parameters $\phi_s$ and $\phi_d$ simultaneously at the Heisenberg-scaling precision, we evaluate the total Fisher information matrix $F=F^\mathcal{S}+F^\mathcal{N}$ in Eqs.~\eqref{EqnC12} and~\eqref{eq:APXCbothFIM}, asymptotically for large $N$, reads
\begin{equation}
F=\frac{4N_sN_c}{16 k^2+1}\left(
\begin{array}{cc}
\beta+\frac{32N_sk^2}{N_c(16k^2+1)} & \sqrt{\beta (1-\beta)} \\
  \sqrt{\beta(1-\beta)} & (1-\beta ) \\
\end{array}
\right),
\end{equation}
where $\beta=N_{c1}/N_c$. The CRB in Eq.~\eqref{eq:CRB} implies that
\begin{align}
(\Delta\phi_s)^2&\geq
\frac{1}{\nu} \frac{(16k^2+1)^2}{128N_s^2k^2} ,
   \label{APXCerrphis}
\\
(\Delta\phi_d)^2&\geq\frac{1}{\nu} 
%\frac{F_{ss}}{F_{dd}-F_{sd}F_{ds}},\\
%&=
\frac{\left(16 k^2+1\right) \left(16 k^2 (\beta N_c+2 N_s)+\beta N_c\right)}{128N_cN_s^2 (1-\beta) k^2} ,
%   \end{split}
  \label{APXCerrphid}
\end{align}
gives the Heisenberg scaling in both parameters simultaneously. However,
we see that the total sensitivity given by  $(\Delta\phi_s)^2+(\Delta\phi_d)^2=\mathrm{Tr}[F^{-1}]$ is maximised by considering $\beta=0$, i.e., the case where only a squeezed vacuum and a coherent state are injected into the MZI which makes the complete FIM is a diagonal matrix, can be written using Eqs.~\eqref{EqnC12} and~\eqref{eq:APXCbothFIM} and given in Eq.~\eqref{eq:DiagMatrix}, reads
\begin{equation}
F=\underbrace{\left(
\begin{array}{cc}
0&0\\0&\frac{4N_cN_s}{16k^2+1}
\end{array}
\right)}_{\text{$F^\mathcal{S}$}}+\underbrace{\left(
\begin{array}{cc}
\frac{128N_s^2k^2}{(16k^2+1)^2}&0\\0&0
\end{array}
\right)}_{\text{$F^\mathcal{N}$}}=\left(
\begin{array}{cc}
\frac{128N_s^2k^2}{(16k^2+1)^2} &0 \\
0 &\frac{4 N_sN_c}{16 k^2+1} \\
\end{array}
\right) .
\end{equation}

 \section{Singularity of $F^\mathcal{S}$ in Eq.~\eqref{eq:FIMdisp}} \label{Appendix:Psedoinverse}
 \numberwithin{equation}{section}
\renewcommand{\theequation}{D\arabic{equation}}       
  \setcounter{equation}{0}
The FIM $F^{\mathcal{S}}$ in Eq.~\eqref{eq:APXCfim1}, which is further simplified by setting $k_1=k_2=k$ as in Eq.~\eqref{eq:FIMdisp},  is asymptotically a singular matrix proportional to $(\sqrt{\beta}, \sqrt{1-\beta})^T (\sqrt{\beta}, \sqrt{1-\beta})$. However, it can achieve Heisenberg scaling for a linear combination of parameters $\phi_s$ and $\phi_d$ given by $\Phi= \sqrt{\beta}\phi_s +\sqrt{1-\beta} \phi_d$. It is evident that the CRB inequality, as defined in Eq.~\eqref{eq:CRB} in the main text, is not well defined due to the singularity of FIM. The pseudo-inverse provides a substitute of the inverse of the FIM and allows for the estimation of parameter sensitivity analysis even in the presence of singularity. It can be obtained through the eigendecomposition of $F^\mathcal{S}$ given by
\begin{equation}
F^\mathcal{S}=VDV^T,
\end{equation}
where $D$ is the diagonal matrix and $V$ be the orthogonal matrix composed of the (column) eigenvectors of $F^\mathcal{S}$ and read
\begin{equation}
D=\begin{pmatrix}
\frac{4Nc Ns}{16k^2+1}&0\\0&0
\end{pmatrix} \text{ \ and \ } V=\begin{pmatrix}
\sqrt{\beta}&-\sqrt{1-\beta}\\
\sqrt{1-\beta}&\sqrt{\beta}
\end{pmatrix}.
\end{equation} 
Thus, the pseudo-inverse matrix $(F^\mathcal{S})^+$ will be represented in the form~\cite{stoica2001parameter}
\begin{equation}
(F^\mathcal{S})^+=VD'V^T,
\end{equation}
where $D'$ is obtained from $D$ by replacing each positive diagonal entry by its reciprocal. Therefore, the pseudo-inverse of $F^{\mathcal{S}}$ in Eq.~\eqref{eq:FIMdisp} reads
\begin{equation}
(F^\mathcal{S})^+=\frac{16k^2+1}{4 N_c N_s}\begin{pmatrix}
\beta & \sqrt{\beta(1-\beta)}\\ \sqrt{\beta(1-\beta)}& 1-\beta
\end{pmatrix}.
\label{eq:PseudoIn}
\end{equation}
The CRB on the sensitivity of the parameter $\Phi$, provided that it is linearly independent of the eigenvectors of the kernel of the FIM, can be indeed written as $(\Delta\Phi)^2\geq (J (F^\mathcal{S})^+ J^T)/\nu$, where $J_i=\partial\Phi/\partial\phi_i$ is Jacobian of $\Phi$ with $i=s,d$. The sensitivity in Eq.~\eqref{eq:CRBDispl} for a linear combination of parameter $\Phi$ can be calculated using Eq.~\eqref{eq:PseudoIn}, reads
\begin{equation}
\begin{split}
(\Delta\Phi)^2\geq& \frac{1}{\nu}J (F^\mathcal{S})^+ J^T\\
  &=\frac{1}{\nu}\frac{16k^2+1}{4N_sN_c}\begin{pmatrix}
\sqrt{\beta} & \sqrt{1-\beta}
\end{pmatrix}\begin{pmatrix}
\beta & \sqrt{\beta(1-\beta)}\\ \sqrt{\beta(1-\beta)}& 1-\beta
\end{pmatrix}\begin{pmatrix}
\sqrt{\beta}\\ \sqrt{1-\beta}
\end{pmatrix}\\
&=\frac{1}{\nu}\frac{16k^2+1}{4N_sN_c}.
\end{split}
\label{eq:APXDPsedo}
\end{equation}
  
\section{Maximum Likelihood Estimators} \label{Appendix E}
\numberwithin{equation}{section}
\renewcommand{\theequation}{E\arabic{equation}}       
  \setcounter{equation}{0} 
In this Appendix, we will solve the equation for the maximum likelihood estimators that saturate the CRB in Eq.~\eqref{eq:CRB}. By repeating the measurement $\nu$ times with two-homodyne detections at the output of the MZI we get $\nu$ set of outcomes $\vec{x}_1,...,\vec{x}_{\nu}$. 
We can then evaluate the likelihood function
\begin{equation}
L(\boldsymbol{\phi}|\vec{x}_1,...,\vec{x}_{\nu})=\prod_{i=1}^{\nu}p(\vec{x}_i|\boldsymbol{\phi}),
\label{EqnE1}
\end{equation}
where $p(\vec{x}_i|\boldsymbol{\phi})$ is the probability density function of the joint homodyne measurement in Eq.~\eqref{eq:ProbDist}. 
Then one can find the maximum likelihood estimators $\Tilde{\boldsymbol{\phi}}$ by maximising the likelihood function $L$.
This can be more conveniently done by maximising the log-likelihood function (as log is a monotonic function)
\begin{equation*}
\begin{split}
0&=\nabla_{\boldsymbol{\phi}}\mathrm{log}L(\boldsymbol{\phi}|\vec{x}_1,...,\vec{x}_\nu)\bigg|_{\boldsymbol{\phi}=\Tilde{\boldsymbol{\phi}}_{MLE}}\\
&=\nabla_{\boldsymbol{\phi}}\sum_{i=1}^{\nu}\mathrm{log} p(\vec{x}_i|\boldsymbol{\phi})\bigg|_{\boldsymbol{\phi}=\Tilde{\boldsymbol{\phi}}_{MLE}}\\
&=\bigg[-\frac{\nu}{2}\nabla_{\boldsymbol{\phi}}\mathrm{log}(\mathrm{det[\Sigma]})-\frac{1}{2}\nabla_{\boldsymbol{\phi}}\sum_{i=1}^{\nu}(\vec{x}_i-\vec{\mu})^T\Sigma^{-1}(\vec{x}_i-\vec{\mu}) \bigg]_{\boldsymbol{\phi}=\Tilde{\boldsymbol{\phi}}_{MLE}}\\
&=\bigg[-\frac{\nu}{2}\mathrm{Tr}\big[\Sigma^{-1}\nabla_{\boldsymbol{\phi}}\Sigma\big]-\frac{1}{2}\nabla_{\boldsymbol{\phi}}\sum_{i=1}^{\nu}\mathrm{Tr}\big[\Sigma^{-1}(\vec{x}_i-\vec{\mu})(\vec{x}_i-\vec{\mu})^T\big] \bigg]_{\boldsymbol{\phi}=\Tilde{\boldsymbol{\phi}}_{MLE}}
\end{split}
\end{equation*}
\begin{equation}
=\bigg[(\nabla_{\boldsymbol{\phi}}\vec{\mu})^T\Sigma^{-1}\bigg(\nu\vec{\mu}-\sum_{i=1}^{\nu}\vec{x}_i\bigg)\bigg]_{\boldsymbol{\phi}=\Tilde{\boldsymbol{\phi}}_{MLE}}+\frac{1}{2}\mathrm{Tr}\bigg[\nabla_{\boldsymbol{\phi}}\Sigma^{-1}\bigg(\nu\Sigma-\sum_{i=1}^{\nu}(\vec{x}_i-\vec{\mu})(\vec{x}_i-\vec{\mu})^\mathrm{T} \bigg)\bigg]_{\boldsymbol{\phi}=\Tilde{\boldsymbol{\phi}}_{MLE}}.
\label{EqnE2}
\end{equation}
Notice that Eq.~\eqref{EqnE2} corresponds to two equations, each one including the derivative of one parameter, either $\phi_s$ or $\phi_d$.
The above equation cannot, in general, be solved analytically. However, it is interesting to note that for a two-parameter estimation problem with a squeezed vacuum and a coherent state input, the above equation is further simplified for a large average number of photons. Here we discuss the case for $\beta=0$ which corresponds to a squeezed vacuum and a coherent state injected into the input port of the MZI. Let us consider the terms in Eq.~\eqref{EqnE2} which are asymptotically dominant for large values of $N$. We see from Eqs.~\eqref{EqnC4}, \eqref{EqnC6}, \eqref{EqnC7} and~\eqref{eq:DMu} that the derivatives $\partial_{\phi_d}\vec{\mu}=O(N^{1/2})$, $\partial_{\phi_s}\vec{\mu}=O(N^{-1/2})$, $\partial_{\phi_d}\Sigma^{-1}=O(N)$, and $\partial_{\phi_s}\Sigma^{-1}=O(N^2)$, whereas $\Sigma^{-1}=C/\mathrm{det}[\Sigma]$ is of order $O(N)$,  $\Sigma$ is of order $O(N^0)$ and $\vec{\mu}$ grows as fast as $N^{1/2}$. The first and second terms of Eq.~\eqref{EqnE2} for $\phi_d$ are of the order $O(N^2)$ and $O(N)$ respectively. In this case, one can consider only the first leading term of Eq.~\eqref{EqnE2} for large $N$:
\begin{equation}
0=\bigg[(\nabla_{\boldsymbol{\phi}}\vec{\mu})^T\Sigma^{-1}\bigg(\nu\vec{\mu}-\sum_{i=1}^{\nu}\vec{x}_i\bigg)\bigg]_{\bm{\phi}=\Tilde{\bm{\phi}}_{MLE}}
\end{equation}
which can be simplified as weighted mean of the estimator $\Tilde{\vec{\mu}}=\frac{1}{\nu}\sum_{i=1}^{\nu}\vec{x}_i$ of the mean $\vec{\mu}$. From Eqs.~\eqref{eq:mu} and~\eqref{eq:APXCMZI} using conditions on the local oscillators in Eq.~\eqref{eq:ConditionLO} by choosing $k_1=k_2$ we can write 
\begin{equation}
\frac{{\mu}_1}{{\mu}_2}=\frac{\Tilde{{\mu}}_1}{\Tilde{{\mu}}_2} \implies -\frac{\sin{\frac{\Tilde{\phi}_d}{2}}}{\cos{\frac{\Tilde{\phi}_d}{2}}}=\frac{\Tilde{{\mu}}_1}{\Tilde{{\mu}}_2}
\end{equation}
This yields an analytical expression for the estimator $\Tilde{\phi}_d$ which reads
\begin{equation}
\Tilde{\phi}_{d\, \mathrm{MLE}}=2\arctan\Bigg[\frac{-\Tilde{\mu}_1}{\Tilde{\mu}_2}\Bigg].
\label{eq:MLEphid}
\end{equation}
Moreover for $\phi_s$ the second term of Eq.~\eqref{EqnE2} is of the order $O(N^2)$ which dominates over the first terms that only grows as fast as $N$ for large values of $N$. In this case, Eq.~\eqref{EqnE2} asymptotically becomes
\begin{equation}
0=\frac{1}{2}\mathrm{Tr}\bigg[\nabla_{\boldsymbol{\phi}}\Sigma^{-1}\bigg(\nu\Sigma-\sum_{i=1}^{\nu}(\vec{x}_i-\vec{\mu})(\vec{x}_i-\vec{\mu})^\mathrm{T} \bigg)\bigg]_{\bm{\phi}=\Tilde{\bm{\phi}}_{MLE}}.
\end{equation}
where it is possible to recognise the sample covariance matrix \begin{equation}
{\Tilde{\Sigma}}=\frac{1}{\nu}\sum_{i=1}^{\nu}(\vec{x}_i-\Tilde{\vec{\mu}})(\vec{x}_i-\Tilde{\vec{\mu}})^T,
\label{eq:APXEstCov}
\end{equation}
as estimator of the covariance matrix $\Sigma$. The function in Eq.~\eqref{eq:APXEstCov} can be inverted, and the
maximum-likelihood estimator for $\phi_s$ reads
\begin{equation}
\Tilde{\phi}_{s\, \mathrm{MLE}}=2\theta_1-2\pi+ \arccos{\Bigg(\frac{2\big(p_1\Tilde{\Sigma}_{11}+p_2\Tilde{\Sigma}_{22}+2\sqrt{p_1p_2}\Tilde{\Sigma}_{12}\big)-\cosh{2r}}{\sinh{2r}}\Bigg)},
\label{eq:MLEphis}
\end{equation}
where $\Tilde{\Sigma}_{ij}$ are the entries of the matrix $\Tilde{\Sigma}$ and $p_1=\cos^2{(\Tilde{\phi}_d/2)}$ with $p_1+p_2=1$. We observe that the estimators $\tilde{\phi}_d$ and $\Tilde{\phi}_s$ in Eqs.~\eqref{eq:MLEphid} and~\eqref{eq:MLEphis} are good approximations of the solutions to Eq.~\eqref{EqnE2} even for smaller values of $N$, as shown in Fig.~\ref{Fig2} and Fig.~\ref{Fig3} of the main text.

\twocolumngrid % Tell bibtex which bibliography style to use
\bibliography{mybib} % Tell bibtex which .bib file to use (this one is some example file in TexLive's file tree)
\end{document}